\documentclass[prb,aps,amsfonts,amssymb,amsmath,floats,floatfix,twocolumn,superscriptaddress,unsortedaddress]{revtex4}
\usepackage{graphics}
\usepackage{epsfig,graphicx}

\begin{document}


\author{I.S. Elfimov}
\affiliation{Department of Physics {\rm {\&}} Astronomy, University of British Columbia, Vancouver, British
Columbia, Canada V6T\,1Z1}
\author{G.A. Sawatzky}
\affiliation{Department of Physics {\rm {\&}} Astronomy, University of British Columbia, Vancouver, British Columbia, Canada V6T\,1Z1}
\title{Theory of Fermi-surface pockets and correlation effects in underdoped YBa$_2$Cu$_3$O$_{6.5}$}
\author{A. Damascelli}
\affiliation{Department of Physics {\rm {\&}} Astronomy, University of British Columbia, Vancouver, British
Columbia, Canada V6T\,1Z1}

\begin{abstract}
The detection of quantum oscillations in the electrical resistivity of YBa$_2$Cu$_3$O$_{6.5}$ provides direct
evidence for the existence of Fermi surface pockets in an underdoped cuprate. We present a theoretical study
of the electronic structure of YBa$_2$Cu$_3$O$_{7-\delta}$ (YBCO) aiming at establishing the nature of these
Fermi pockets, i.e. CuO$_2$ plane versus CuO chain or BaO. We argue that electron correlation effects, such
as orbital-dependent band distortions and highly anisotropic self-energy corrections, must be taken into
account in order to properly interpret the quantum oscillation experiments.
\end{abstract}

\date{\today}

\pacs{}

\maketitle

\begin{figure}[b]
\centerline{\epsfig{figure=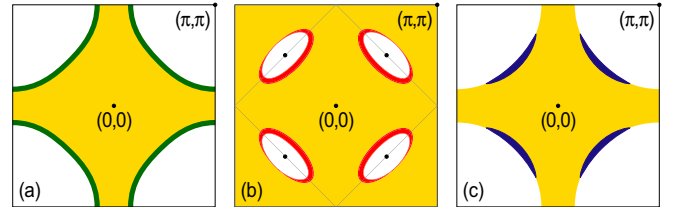,width=1\linewidth,clip=}} \caption{(Color online) Different scenarios for the low-energy
electronic structure of the single, square CuO$_2$ plane; Brillouin zone dimensions are expressed modulo the lattice constant. (a)
Independent-particle Fermi surface for a metallic, 1/2-filled Cu-O band (green/gray); its volume is precisely 50\% of the zone. When
correlations are considered, the system becomes an antiferromagnetic insulator; (b) light hole-doping might result in four Fermi pockets
(red/gray) with a volume that, counting holes, corresponds to the doping $x$ away from 1/2-filling (the enclosed diamond is the
antiferromagnetic Brillouin zone). (c) Alternatively, the Fermi surface is truncated near the zone edges reducing to four disconnected arcs
(blue/gray). In all panels, white and yellow (gray) areas identify hole and electron momentum-space regions, respectively; the thickness of the
Fermi surface lines represents the quasiparticle coherence $Z_{\bf k}$ (i.e., the spectral weight of the electron-removal quasiparticle peaks),
which is strongly $k$-dependent in (b) and (c).} \label{fig_1}
\end{figure}

\begin{figure*}[t]
\centerline{\epsfig{figure=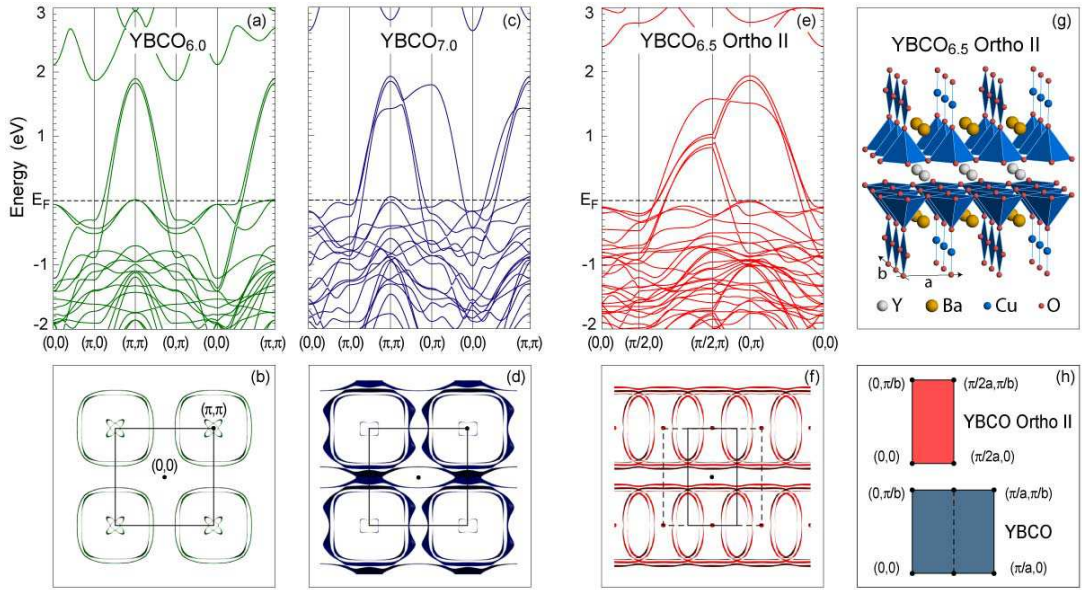,width=0.8\linewidth,clip=}} \vspace{-0.2cm} \caption{(Color online) Local-density
approximation (LDA) calculations of the in-plane electronic structure ($k_z\!=\!0$), and corresponding two-dimensional projected Fermi surface,
for (a,b) YBCO$_{6.0}$, (c,d) YBCO$_{7.0}$, and (e,f) oxygen-ordered ortho-II YBCO$_{6.5}$ (the width of the Fermi surface lines is proportional
to the amount of $k_z$ dispersion). (g) Ortho-II YBCO$_{6.5}$ crystal structure, with full and empty CuO chains alternating along the $a$-axis
and running parallel to the $b$-axis (plane-octahedra and chain-plaquettes each contain one Cu atom). (h) Sketch of 1/4$-$Brillouin zone for
ordinary YBCO and ortho-II YBCO$_{6.5}$; because of the folding to the reduced zone, ($\pi/a$,0)-(0,0) and ($\pi/a$,$\pi/b$)-(0,$\pi/b$) are
sets of equivalent symmetry points in YBCO$_{6.5}$. In (a-f), as throughout the paper, momentum-space coordinates are given modulo the lattice
parameters $a$ and $b$.} \label{fig_2}
\end{figure*}

The independent-particle picture fails miserably in the description
 of the undoped parent compound of the HTSCs, such as for instance
La$_2$CuO$_4$ or YBCO$_{6.0}$. These systems, as shown schematically in Fig.\,\ref{fig_1}a for a single
square CuO$_2$ plane or in Fig\,\ref{fig_2}a,b for YBCO$_{6.0}$, are predicted by band theory to be
1/2-filled metals; instead, due to strong electronic correlations, they are charge-transfer gap
antiferromagnetic insulators\cite{george,dagotto}. While on the overdoped side of the phase diagram a
well-defined band-structure-like Fermi surface is recovered, as in the case of heavily overdoped
Tl$_2$Ba$_2$CuO$_{6+\delta}$\cite{Hussey:2003,29a,darren}, the underdoped regime is the most debated, with
two alternative scenarios being proposed. The first originates from a wide variety of theoretical
calculations\cite{kampf,wen,sushkov}, which predict that the first few holes doped in to the antiferromagnet
would give rise to four `nodal Fermi pockets' with a volume, counting holes, proportional to the doping $x$
away from half filling (Fig.\,\ref{fig_1}b). The second consists of the truncation of the single-particle
Fermi surface giving rise to four `nodal Fermi arcs' (Fig.\,\ref{fig_1}c); it is suggested based on
experimental observations by angle-resolved photoemission spectroscopy (ARPES)\cite{norman,kanigel} and is
naturally connected to the existence of a $d$-wave-like pseudogap\cite{loeserPG,dingPG,kmshen,stefan}.

The discrimination between the two scenarios of Fig.\,\ref{fig_1}b,c has mainly relied on ARPES studies; one should realize, however, that this
is a formidable and yet unsettled task\cite{andrea,campuzano}. It rests on the detection of an arc profile, or lack thereof, on the outer side
of the pockets, where the electron-removal spectral weight is expected to be vanishingly small because of the strong drop in the quasiparticle
coherence $Z_{\bf k}$ beyond the antiferromagnetic zone boundary (diamond in Fig.\,\ref{fig_1}b)\cite{eder}.
The discovery of a closed Fermi surface contour in quantum oscillation experiments on ortho-II
YBCO$_{6.5}$\cite{Louis2007} might lead to a resolution of this longstanding conundrum, by providing direct
support to the nodal Fermi pocket scenario of Fig.\,\ref{fig_1}b. For this assignment to be conclusive,
however, one must exclude the possibility that bands other than the Cu$(3d_{x^2-y^2})$-O$(2p_{x,y})$ from the
CuO$_2$ planes, such as for instance the BaO bands, are responsible for the detected signal. Note in this
regard that the BiO and TlO pockets predicted for instance for Bi\cite{bi2212LDA} and Tl\cite{tl2201LDA}
cuprates have never been detected experimentally; on the other hand, the situation for YBCO is much more
complicated because of the presence of the CuO chains, which can potentially give rise to Fermi surface
sheets of purely chain character and provide an additional channel for BaO hybridization. The presence of
such BaO/CuO-chain based Fermi pockets in band theory calculations for optimally doped YBCO has been pointed
out previously by Andersen et al.\cite{DFT_YBCO_literature1}.

To illustrate the band structure effects on the Fermi surface of the YBCO family, we have carried out density
functional theory calculations of the electronic structure of YBCO$_{7-\delta}$, for $\delta\!=\!0$, 0.5, and
1 (Fig.\,\ref{fig_2} and Ref.\,\onlinecite{method}). In agreement with previous density functional
studies\cite{DFT_YBCO_literature1,DFT_YBCO_literature2}, the Fermi surface of stoichiometric YBCO$_{6.0}$ and
YBCO$_{7.0}$ consists of two large cylindrical sheets, originating from the CuO$_2$ bilayers, and small
pockets of mainly BaO character (see below), all centered around the Brillouin zone corner
(Fig.\,\ref{fig_2}a-d). In addition, in YBCO$_{7.0}$ there is also the open one-dimensional Fermi surface of
the `full' CuO chains running along the $b$-axis (Fig.\,\ref{fig_2}d). In oxygen ordered ortho-II
YBCO$_{6.5}$, the unit cell is doubled along $a$-axis due to the alternation of full and empty CuO chains
and, correspondingly, the Brillouin zone is folded along the ($\pi/2,0)\!-\!(\pi$/2,$\pi$) line
(Fig.\,\ref{fig_2}g,h; momentum-space coordinates are here given modulo the lattice constants). This affects
both shape and number of Fermi surface sheets, but the small BaO pockets are still present although
translated to (0,$\pm$$\pi$).

As one can see in Fig.\,\ref{fig_2}f, if taken quantitatively the size of the (0,$\pm$$\pi$) pockets is
vanishingly small and is not nearly enough to explain the quantum oscillations in the electrical resistance
observed by Doiron-Leyraud et al.\cite{Louis2007}. However, it is also evident that in such a multiband case
the Fermi surface is quite susceptible to very small changes of the relative position of the bands and/or to
a minimal rigid shift of the Fermi level itself, which can be induced by deviations in oxygen concentration
or the presence of disorder.
\begin{figure}[b!]
\vspace{-0.25cm}
\centerline{\epsfig{figure=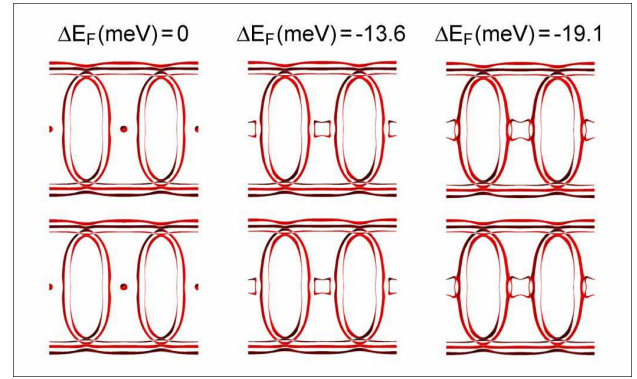,width=.95 \linewidth,clip=}} \vspace{-0.25cm} \caption{(Color online) LDA Fermi surface
of ortho-II YBCO$_{6.5}$ for different positions of the Fermi level, so as to show the rigid-band-like doping dependence of Fermi surface and,
in particular, BaO-Cu$_{chain}$ pockets at (0,$\pi$) and ($\pi$,$\pi$).} \label{fig_3}
\end{figure}
For instance (Fig\,\ref{fig_3}), for Fermi level position $\Delta E_F\!=\!0$, $-13.6$, and $-\!19.1$\,meV the area of the (0,$\pm$$\pi$) BaO
pocket is $A_k\!=\!0.3$, 3.01, and 5.22\,nm$^{-2}$ respectively, to be compared to $A_k\!=\!5.1$\,nm$^{-2}$ found in quantum oscillation
experiments on YBCO$_{6.5}$\cite{Louis2007}. Also the BaO pocket average band mass, $m_b\!=\!0.77$, 1.68, and 1.56\,$m_0$ for the three values
of $\Delta E_F$, is in qualitative agreement with the experimentally observed cyclotron mass $m^{\star}\!=\!1.9\,m_0$, where $m_0$ is the free
electron mass\cite{Louis2007}.

\begin{figure}[t!]
\centerline{\epsfig{figure=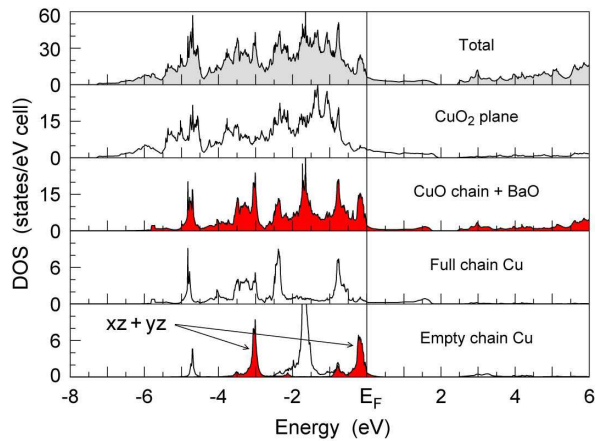,width=0.9\linewidth,clip=}} \vspace{-0.25cm} \caption{(Color online) LDA density of
states (DOS) for ortho-II YBCO$_{6.5}$. The high DOS at $E_F$ originates from BaO and (oxygen-empty) Cu$_{chain}$ orbitals; these are $pd\pi$
orbitals with Cu $d_{xz,yz}$ character, thus very different from the $pd\sigma$ orbitals typically found for the CuO$_2$ planes ($d_{x^2-y^2}$)
or that one might expect for the empty CuO chains ($d_{3z^2-r^2}$)\cite{DFT_YBCO_literature1}.} \label{fig_4}
\end{figure}

Before elaborating on the deeper significance of these findings, we would like to discuss the anticipated BaO
assignment of the small  Fermi surface pockets. As shown in Fig.\,\ref{fig_4} for YBCO$_{6.5}$, a strong peak
in the density of states (DOS) at energies just below the Fermi level belongs to Cu $3d_{xy}$ and $3d_{yz}$
orbitals which are, in fact, strongly mixed with the apex O $2p$ and Ba empty orbitals in a $pd\pi$
configuration\cite{DFT_YBCO_literature1}. Thus, BaO-Cu$_{chain}$ is primarily the character of the flat bands
(Fig.\,\ref{fig_2}e) and corresponding small pockets (Fig.\,\ref{fig_2}f) about (0,$\pm$$\pi$) in
YBCO$_{6.5}$.
A similar analysis for YBCO$_{6.0}$ confirms the same origin for the ($\pm\pi$,$\pm\pi$) pockets; as for
YBCO$_{7.0}$, which is somewhat of a different case because the CuO chains are all full, we find that oxygen
from the BaO plane contributes the most to those bands defining the small hole pockets. Consistent with our
assignment, all these pockets are rather isotropic in $k_x$ and $k_y$, contrary to what is expected for a
purely one-dimensional chain electronic structure. Also, they are very sensitive to the
Cu$_{chain}$-O$_{apex}$ and -Ba distances, which show a substantial variation with $\delta$ in
YBCO$_{7-\delta}$\cite{structure}.

\begin{figure}[t!]
\centerline{\epsfig{figure=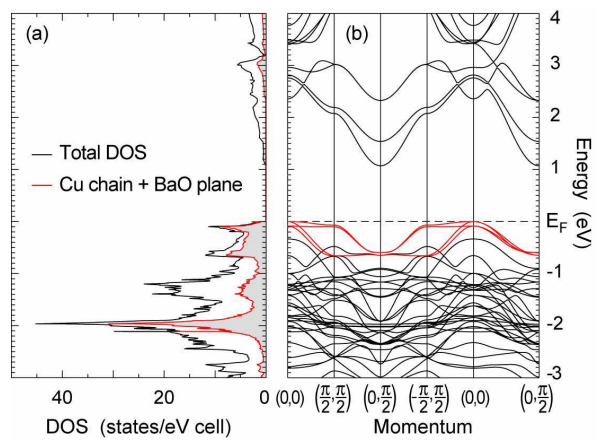,width=.9\linewidth,clip=}} \vspace{-0.25cm} \caption{(Color online) Density of states
(DOS) and band dispersion calculated for YBCO$_{6.0}$ within the LDA+U approximation, with parameters $U\!=\!8$\,eV and $J_H\!=\!1.34$\,eV
\cite{method,parameters}. The system is now a 1.08\,eV gap charge-transfer insulator, with a Fermi level DOS dominated by the BaO-Cu$_{chain}$
bands; the CuO$_2$ plane band is split and pushed to high energies.} \label{fig_5}
\end{figure}

At this point the key question is: what is the relevance of the results we have just presented to the physics
of the underdoped cuprates?  After all, these are strongly correlated electron systems and it is well-known
that the standard local-density approximation (LDA) to the exchange correlation potential fails to treat
strong correlations properly. In order to gain more direct insight into this very fundamental methodological
shortcoming, we further our analysis within the so-called LDA+U method \cite{method} which, by adding the
on-site Coulomb repulsion U to LDA in a mean field-like way, can reproduce the correct magnetic ground state
and forbidden gap in correlated antiferromagnetic insulators such as La$_2$CuO$_4$\cite{george,ldau}.
However, the LDA+U approach works fairly well only in the case of integer orbital occupation and
well-established long-range order (e.g., spin, charge, or orbital order) \cite{george2}. This constrains our
DFT analysis of the Fermi surface of YBCO in the presence of strong correlations to the single case of
YBCO$_{6.0}$ (Fig.\,\ref{fig_5}), which is truly an antiferromagnetic insulator with $3d^9$ and $3d^{10}$
configurations for plane-Cu$^{2+}$ and chain-Cu$^{1+}$, respectively. It is important to realize that in such
a framework, as shown in Fig.\,\ref{fig_5}, the CuO$_2$ in-plane states are almost completely removed from
the Fermi energy and the first electron-removal state has nearly 100\% BaO-Cu$_{chain}$ character. This is
partly due to the inadequacy of LDA+U in fully accounting for the many-body physics of the CuO$_2$ planes;
while the LDA in-plane states are about 0.5\,eV below the Fermi level, we do know that in the undoped
cuprates there are Zhang-Rice singlet (ZRS) states\cite{zhang} pushed $\sim\!0.5$\,eV higher in energy, thus
back into the first ionization energy range, but these two-particle states are completely missed by the
single-particle LDA description\cite{george}.
In the scope of this paper, we can make one very important observation: independently of the exact energy of
the ZRS states, in YBCO$_{6.0}$ a rigid-band-like hole doping would give rise also to pockets originating
from the BaO-Cu$_{chain}$ bands and, in direct contrast to materials like La$_2$CuO$_4$, not just the
in-plane Cu$(3d_{x^2-y^2})$-O$(2p_{x,y})$ states. Furthermore, the folding of the Brillouin zone, in this
case due to antiferromagnetic spin order, translates the BaO-Cu$_{chain}$ pocket from ($\pi$,$\pi$) to the
$\Gamma$-point (Fig.\,\ref{fig_1}a and\,\ref{fig_5}b).

Let us now attempt a critical summary of experimental and theoretical findings. In light of our understanding
of the underdoped cuprates, it is definitely very tempting to interpret the recent quantum oscillations
results\cite{Louis2007} as direct evidence for nodal Fermi pockets originating from the CuO$_2$ planes
electronic states. Our analysis shows, however, that the details of the band structure cannot be neglected in
the case of YBCO$_{6.5}$, since BaO-Cu$_{chain}$ states might give rise to additional Fermi pockets
consistent with the recent experimental observations both as far as volume and especially small effective
mass are concerned. As a matter of fact, their role might become even more significant when correlation
effects are also included, as they should for these very low doping values. Although not as extreme as in the
case of YBCO$_{6.0}$, the Coulomb repulsion U associated with the Cu 3$d$ orbitals would strongly affect the
in-plane CuO$_2$ bands but much less so the BaO-Cu$_{chain}$ bands. This would lead to an orbital-dependent
band distortion (as opposed to a rigid shift) and trigger a hole transfer from the more to the less
correlated electronic states, conspiring to generate qualitatively different Fermi features:
`highly-correlated' CuO$_2$ pockets playing a primary role in the emergence of high-temperature
superconductivity, and `more conventional' BaO-Cu$_{chain}$ pockets behaving as mere spectators. In the event
of such coexistence, the latter would be more likely observed in quantum oscillation experiments, while the
CuO$_2$ pockets might escape detection because of the highly momentum-dependent self-energy effects, which
could even result in vanishing quasiparticle residues and/or diverging effective masses at some momenta along
a Fermi pocket contour (e.g., beyond the antiferromagnetic zone boundary as in Fig.\,\ref{fig_1}b). If this
was indeed the case, it would also suggest that the detection of quantum oscillations might be limited to
those cuprates containing CuO chains \cite{124}.

Due to the complex multiband and correlated character of the electronic structure of YBCO$_{6.5}$, the
precise determination of the nature of the Fermi surface pockets revealed by quantum oscillation
experiments\cite{Louis2007} will require further study of this underdoped system by a variety of techniques.
One serious issue, which has not been addressed here, concerns the negative sign of the Hall coefficient
observed by Doiron-Leyraud et al.\cite{Louis2007}. In a simple one particle picture, this negative sign would
point to an electron rather than hole-like character for the Fermi surface pockets detected in quantum
oscillations. As in Ref.\,\onlinecite{Louis2007},  we also concentrated on the oscillatory part of the Hall
resistivity above and beyond the rather smooth ``background'' contribution from which the negative sign is
derived. We note that this Hall resistivity background is expected to be rather complicated because of
contributions from the strongly correlated CuO$_2$ planes and the presence of localized magnetic moments; a
net negative sign for the Hall coefficient need not be inconsistent with the presence of hole-like Fermi
surface pockets.

{\it Acknowledgments:} This work was supported by the Alfred P. Sloan Foundation (A.D.), the CRC Program
(A.D. and G.A.S), NSERC, CFI, CIFAR, and BCSI.

{\it Note:} After completion of this work, we became aware of a recent DFT study of YBCO by Carrington and
Yelland \cite{carrington}. The LDA band structure results are very similar, with minor exceptions due to the
difference in the treatment of the exchange correlation potential\cite{carrington}.

\bibliographystyle{plain}

\begin{thebibliography}{99}
\bibitem{george}J. Zaanen, G.A. Sawatzky, and J. W. Allen, Phys. Rev. Lett. {\bf 55}, 418 (1985);
M.T. Czyzyk and G.A. Sawatzky Phys. Rev. B {\bf 49}, 14211 (1994).
\bibitem{dagotto}E. Dagotto, Rev. Mod. Phys. {\bf 66}, 763 (1994).
\bibitem{Hussey:2003}N.E. Hussey, M. Abdel-Jawad, A. Carrington, A.P. Mackenzie, and L. Balicas, Nature {\bf 425}, 814 (2003).
\bibitem{29a}M. Plat$\acute{{\rm e}}$, J.D.F. Mottershead, I.S. Elfimov, D.C. Peets, R. Liang, D.A. Bonn, W.N. Hardy,
S. Chiuzbaian, M. Falub, M. Shi, L. Patthey, and A. Damascelli, Phys. Rev. Lett. {\bf 95}, 077001 (2005).
\bibitem{darren}D.C. Peets, J.D.F Mottershead, B. Wu, I.S. Elfimov, R. Liang, W.N. Hardy, D.A. Bonn, M. Raudsepp,
N.J.C. Ingle, and A. Damascelli, New J. Phys. {\bf 9}, 1 (2007).
\bibitem{kampf}A.P. Kampf and J.R. Schrieffer, Phys. Rev. B {\bf 42}, 7967 (1990).
\bibitem{wen}X.G. Wen and P.A. Lee, Phys. Rev. Lett. {\bf 76}, 503 (1996).
\bibitem{sushkov}O.P. Sushkov, G.A. Sawatzky, R. Eder, and H. Eskes, Phys. Rev. B {\bf 56}, 11769 (1997).
\bibitem{norman}M.R. Norman, H. Ding, M. Randeria, J.C. Campuzano, T. Yokoya, T. Takeuchi, T. Takahashi, T. Mochiku,
K. Kadowaki, P. Guptasarma, and D.G. Hinks, Nature {\bf 392}, 157 (1998).
\bibitem{kanigel}A. Kanigel, M.R. Norman, M. Randeria, U. Chatterjee, S. Souma, A. Kaminski, H.M. Fretwell, S. Rosenkranz,
M. Shi, T. Sato, T. Takahashi, Z.Z. Li, H. Raffy, K. Kadowaki, D. Hinks, L. Ozyuzer, and J.C. Campuzano,
Nature Physics {\bf 2}, 447 (2006).
\bibitem{loeserPG}A.G. Loeser, Z.-X. Shen, D.S. Dessau, D.S. Marshall, C.H. Park, P. Fournier, and A. Kapitulnik,
Science {\bf 273}, 325  (1996).
\bibitem{dingPG}H. Ding, T. Yokoya, J.C. Campuzano, T. Takahashi, M. Randeria, M.R. Norman, T. Mochiku, K. Hadowaki,
and J. Giapintzakis, Nature {\bf 382}, 51 (1996).
\bibitem{kmshen}K.M. Shen, F. Ronning, D.H. Lu, F. Baumberger, N.J.C. Ingle, W.S. Lee, W. Meevasana, Y. Kohsaka, M. Azuma,
M. Takano, H. Takagi, and Z.-X. Shen, Science {\bf 307}, 901  (2005).
\bibitem{stefan}S. H\"{u}fner, M.A. Hossain, A. Damascelli, and G.A. Sawatzky, arXiv:0706.4282 (2007).
\bibitem{andrea}A. Damascelli, Z. Hussain, and Z.-X. Shen, Rev. Mod. Phys. {\bf 75}, 473 (2003).
\bibitem{campuzano}J.C. Campuzano,  M.R. Norman, and M. Randeria, {\it Photoemission\,in\,the\,high-$T_c$ superconductors}.
In {\it The Physics of Superconductors} Vol.\,II\,(eds K.H.\,Bennemann and J.B.\,Ketterson)\,167\,(Springer,
2004).
\bibitem{eder}H. Eskes and R. Eder, Phys. Rev. B {\bf 54}, 14226 (1996).
\bibitem{Louis2007}N. Doiron-Leyraud, C. Proust, D. LeBoeuf, J. Levallois, J.-B. Bonnemaison, R. Liang, D.A. Bonn,
W.N. Hardy, and L. Taillefer, Nature {\bf 447}, 565 (2007).
\bibitem{bi2212LDA}H. Lin, S. Sahrakorpi, R.S. Markiewicz, and A. Bansil, Phys. Rev. Lett. {\bf 96}, 097001 (2006).
\bibitem{tl2201LDA}S. Sahrakorpi, H. Lin, R.S. Markiewicz, and A. Bansil, Physica C {\bf 460}, 428 (2007).
\bibitem{DFT_YBCO_literature1}O.K. Andersen, O. Jepsen, A.I. Liechtenstein, and I.I. Mazin,
Phys. Rev. B {\bf 49}, 4145 (1994); O.K. Andersen, A.I. Liechtenstein, O. Jepsen, and F. Paulsen, J. Phys.
Chem. Solids {\bf 56}, 1573 (1995).
\bibitem{method} The YBCO$_{7-\delta}$ band structure, for $\delta\!=\!0$,\,0.5, and\,1, was calculated with the
full-potential linearized augmented plane-wave density functional theory (DFT) code WIEN2K. The calculations
for $\delta=$\,0 and 1 were performed using the crystal structures reported by Jorgensen et
al.\cite{jorgensen} for $\delta$=0.07 and 0.91, respectively, while the $\delta\!=\!0.5$ structure was taken
from the paper by Grybos et al.\cite{grybos}. This is because the total energy for the latter is 0.22\,eV
lower than for the structure refined by Jorgensen et al. for a similar oxygen concentration. All calculations
were done with fixed a basis set and muffin tin radii: $R_Y$=2.25, $R_{Ba}$=2.45, $R_{Cu}$=1.76 and
$R_O$=1.56 a.u. The exchange and correlation effects are treated within the local density approximation
(LDA), after Perdew and Wang\cite{lda}. LDA+U calculations for YBCO$_{6.0}$ were performed with $U\!=\!8$\,eV
and $J_H\!=\!1.34$\,eV for Cu 3$d$ Coulomb repulsion and  Hund's coupling, as reported in the literature
\cite{parameters}. We have also performed calculations for other values of $U$; no qualitative changes were
observed down to $U\!=\!4$\,eV, with the main discernible effect being the reduction of the gap upon
decreasing $U$.
\bibitem{jorgensen}J.D. Jorgensen, B.W. Veal, A.P. Paulikas, L.J. Nowicki, G.W. Crabtree, H. Claus,
and W.K. Kwok, Phys. Rev. B {\bf 41}, 1863 (1990).
\bibitem{grybos}J. Grybos, D. Hohlwein, Th. Zeiske, R. Sonntag, F. Kubanek, K. Eichhorn, and Th. Wolf,
Physica C {\bf 220}, 138 (1994).
\bibitem{lda}J.P. Perdew and Y. Wang, Phys. Rev. B {\bf 45}, 13244 (1992).
\bibitem{DFT_YBCO_literature2}E. Bascones, T.M. Rice, A.O. Shorikov, A.V. Lukoyanov, and V.I. Anisimov,
Phys. Rev. B {\bf 71}, 012505 (2005).
\bibitem{structure} The Cu$_{chain}$-Ba distance in YBCO$_{6.0}$ is 3.57\,\AA\ while is 3.47\,\AA\ in
YBCO$_{7.0}$\cite{jorgensen}; on the other hand, the Cu-O$_{apex}$ bond increases from 1.8 to 1.86\,\AA.
Total-energy calculations show that a shift of this magnitude along the $c$-axis of Ba (apex O) towards (away
from) the chains in YBCO$_{6.0}$ results in the lowering of the energy of the BaO-Cu$_{chain}$ states and, in
turn, in the disappearance of the corresponding Fermi surface pockets. At the same time, however, this
structural distortion leads to an increase in the total energy of 0.35eV, which makes it rather unphysical.
\bibitem{ldau} V.I.~Anisimov, J. Zaanen, and O.K. Andersen, Phys. Rev. B {\bf 44}, 943 (1991)
\bibitem{george2}http://online.itp.ucsb.edu/online/motterials07/sawatzky/
\bibitem{zhang}F.C. Zhang and T.M. Rice, Phys. Rev. B {\bf 37}, 3759 (1988).
\bibitem{parameters}J. Ghijsen, L.H. Tjeng, J. van Elp, H. Eskes, J. Westerink, G.A. Sawatzky, and M.T. Czyzyk,
Phys. Rev. B {\bf 38}, 11322 (1988); L.H. Tjeng, C.T. Chen, and S-W. Cheong, Phys. Rev. B {\bf 45}, 8205
(1992).
\bibitem{124}Quantum oscillations were observed also in double-chain
YBa$_2$Cu$_4$O$_8$, for which standard band theory predicts a lack of BaO-Cu$_{chain}$ Fermi surface pockets
\cite{yelland,draxl,yu}. As discussed in Ref.\,\onlinecite{structure}, the relative energy of those bands is
very sensitive to lattice distortions, in addition to correlation effects; at this stage it is not clear
whether such distortions are present in this compound. More detailed experimental and theoretical studies are
needed to clarify this issue.
\bibitem{yelland}E.A. Yelland, J. Singleton, C.H. Mielke, N. Harrison, F.F. Balakirev, B. Dabrowski, and J.R. Cooper,
arXiv:0707.0057 (2007).
\bibitem{draxl}C. Ambrosch-Draxl, P. Blaha, and K. Schwarz, Phys. Rev. B {\bf 44}, 5141 (1991).
\bibitem{yu}J. Yu, K.T. Park, and A.J. Freeman, Physica C {\bf 172}, 467 (1991).
\bibitem{carrington}A.~ Carrington and E.A.~Yelland, Phys. Rev. B {\bf 76}, 140508 (2007).
\end{thebibliography}

\end{document}